# Title

Can ultraprecision polishing techniques improve the coherence times of nitrogen-vacancy centers in diamond?

# Authors

G. Braunbeck[1], S. Mandal[2], M. Touge[3], O. A. Williams[2] and F. Reinhard[1,*]

[1] Walter Schottky Institut and Physik Department, Technische Universität München, Am Coulombwall 4, 85748 Garching, Germany

[2] School of Physics and Astronomy, Cardiff University, Queen's Buildings, The Parade, Cardiff, CF24 3AA, UK

[3] Kumamoto University, Kurokami 2-39-1, Chuou-ku, Kumamoto, 860-8555, Japan

*friedemann.reinhard@wsi.tum.de

# Abstract

We investigate the correlation between surface roughness and corresponding $T_2$ times of near-surface nitrogen-vacancy centers (~7 nm/ 5 keV implantation energy) in diamond. For this purpose we compare five different polishing techniques, including both purely mechanical as well as chemical mechanical approaches, two different substrate sources (Diam2tec and Element Six) and two different surface terminations (O- and H-termination) during nitrogen-vacancy forming. All coherence times are measured and compared before and after an oxygen surface treatment at 520 °C.

We find that the coherence times of shallow nitrogen-vacancy centers are surprisingly independent of surface roughness.

# Introduction

Quantum nanosensing has recently become a very promising application of diamond devices due to a specific quantum sensor: the nitrogen-vacancy (NV) center. A single center embedded few nanometers beneath the diamond surface is sufficiently sensitive to record nuclear magnetic resonance spectra of nanoscale samples[1,2]. This technique of "nanoscale NMR" has led to numerous applications including the conformation analysis of ice[3], as well as the detection and spectroscopy of atomically thin layers[4] or even single proteins[5]. Simultaneously, NV centers in scanning probes have enabled imaging of antiferromagnetic domains[6] and nanodiamonds containing NV defects have enabled temperature measurements in living cells[7,8]. As many signals decay steeply with the sensor-sample distance (cubically for a single spin), NV centers have to be located very close to the diamond surface for sensing applications. Although it is possible to implant optically stable NVs at a depth of only 1.1 nm, the resulting electron spin $T_2$ times are reduced more than tenfold compared to NVs in the bulk[9,10] for reasons that are not fully understood to date[11,12].

As a consequence, much effort has been spent on improving coherence by better surface preparation: thermal oxidation, as will be investigated in our study, can improve the $T_2$ time, although its impact varies considerably (improvement of effective $T_2^*$ from ~10 µs to ~50 µs in ref [5], ~15 µs to ~18 µs in ref [13]). Oxidative etching[9,18] and plasma etching[19,20] seem to reliably create shallow NVs from deeper implantations while maintaining relatively good coherence times. While O-, H- and F-terminations have been tested without success[21], promising calculations on N-terminations[22–24], as also considering different surface orientations of the diamond[24], could not be confirmed experimentally yet. However, simply covering the diamond with glycerol is reported to increase $T_2$ from 28 µs to 132 µs[12*]. Complementary approaches aim to influence the coherence properties during or before NV center creation: Shallow NVs created in a delta-doping process can have coherence times $T_2 > 100$ µs[25], although this result is not easily repeatable[15,26]. Growing a thin layer of $SiO_2$ before implanting the nitrogen ions does not increase the coherence time[27]. However, growing a thin layer of boron-doped diamond before vacuum annealing and removing it afterwards has been reported to improve $T_2$ up to 180 µs for shallow NV centers[28].

---

[*] to ease comparison, all $T_2$ times in this paragraph have been corrected for their implantation depth and dynamical decoupling[14] order, assuming a $d^2$ rise of $T_2$ with surface distance $d$ [15] and a $N^{1/2}$ scaling with the number of decoupling pulses $N$ [15–17].

Intuitively, the surface morphology is another parameter which may affect the local electric[12] and magnetic[16,21] field noise, which are likely origins of the observed decoherence. So far, no systematic study of influence of the surface geometry has been reported. It is known that different polishing techniques can yield surface roughnesses varying by two orders of magnitude[29], with the best results approaching smoothness on the atomic level[30–32].

We therefore investigated the influence of different polishes on the coherence time $T_2$ of shallow implanted NVs. We applied different polishing methods to a set of 10 diamonds, subsequently implanted NV centers, and performed Hahn echo[33] measurements before and after a 520 °C thermal oxidation step (Figure 1).

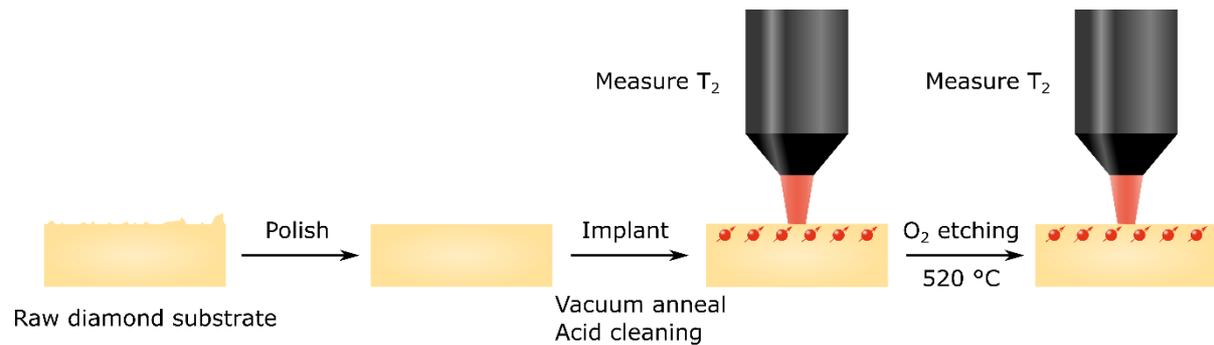

*Figure 1: Investigating the correlation between surface geometry and the coherence time $T_2$. A set of diamonds is subjected to different polishing techniques and subsequently doped with NV centers by ion implantation, vacuum annealing and acid cleaning. For each diamond, the NV spin coherence time $T_2$ is measured before and after a 520 °C thermal oxidation step. (2 columns)*

## Material and Methods

One electronic grade single-crystal CVD diamond with a scaife-polished (100) surface and dimensions of 2.0 x 2.0 x 0.3 mm³ was bought from Diam2tec GmbH, referred to as substrate 1. We measured a surface roughness of *rms* = 2.07 nm by AFM with clearly visible polishing grooves (Fig. 2b), consistent with the manufacturer's specification of *rms* < 5.5 nm[†]. We will refer to this sample as sample Std3. All other diamonds investigated have been laser-cut from three single-crystal electronic grade CVD diamonds with a scaife-polished (100) surface from Element Six Ltd (145-500-0389), referred to as substrates 2A through 2C. Since our study focuses on the effect of surface roughness, no additional treatment was applied to reduce potential subsurface defects.

Two diamonds (referred to as Std1 and Std2) were chosen to keep the manufacturer's original scaife polish, with an rms roughness of roughly 1 nm (see table 1 for details) and clearly visible polishing grooves in AFM (Fig 2a).

---

* to ease comparison, all $T_2$ times in this paragraph have been corrected for their implantation depth and dynamical decoupling[14] order, assuming a $d^2$ rise of $T_2$ with surface distance $d$ [15] and a $N^{1/2}$ scaling with the number of decoupling pulses $N$ [15–17].

† *rms* value estimated from $R_a$ value via the following equation: *rms* ≈ *1.1 $R_a$*

| Sample name | Substrate | Polishing technique | Termination | Average roughness (rms) |
|---|---|---|---|---|
| **Std1** | 2A | - | O (acid clean) | 1.29 nm |
| **Std2** | 2C | - | O (acid clean) | 0.59 nm |
| **Scf1** | 2B | Scaife | O (acid clean) | 2.42 nm |
| **Scf2** | 2C | Scaife | O (acid clean) | 1.07 nm |
| **Cmp1** | 2A | CMP | O (acid clean) | 2.68 nm |
| **Cmp2** | 2C | CMP | O (acid clean) | 0.51 nm |
| **Uv1** | 2A | UV-CMP | O (acid clean) | 0.44 nm |
| **Uv2** | 2B | UV-CMP | O (acid clean) | 0.56 nm |
| **H1** | 2A | - | H (plasma) | 0.95 nm |
| **Std3** | 1 | - | O (acid clean) | 2.07 nm |

*Table 1: Overview of substrate, preparation and resulting roughness for each sample of this study.*

Two samples (Scf1 and Scf2) were subjected to a different scaife polish (Almax easyLab), resulting in surface roughnesses of *rms* = 2.42 nm (Scf1) and 1.07 nm (Scf2, Figure 2f) with visible polishing grooves.

One diamond (H1) was treated by hydrogen plasma (15min in the Seki Technotron quartz tube reactor of an ASTEX microwave plasma system, power 750 W, pressure 50 mbar, flow 100 sccm, temperature 700 °C). While this had no effect on roughness (table 1), termination-induced lattice charging during annealing could affect the NV creation process[28].

Two samples (Cmp1 and Cmp2) were processed in a recently developed chemical mechanical polishing (CMP) process employing a fluid of silica nanoparticles[30,34]. On these samples we observe point-like defects rather than grooves, and one of them shows a significantly reduced roughness of *rms* = 0.51 nm (Cmp2, Figure 2d). This latter result varies, with Cmp1 showing roughness of *rms* = 2.68 nm, presumably owing to local variations in substrate properties.

Finally, two samples (UV1 and UV2) were polished by another recently developed CMP process, based on a UV-light-induced photochemical reaction[31]. While there are comparably high debris-like defects on both resulting surfaces, they are extremely smooth for the most part. The clean areas provide roughnesses of *rms* = 0.44 nm (Uv1) and *rms* = 0.56 nm (Uv2, Figure 2e).

After polishing, all diamonds were implanted with $^{15}N^+$-ions with an implantation energy of 5 keV and an ion flux of $10^{10}$ ions/cm² under an angle of 7° (CuttingEdge Ions, LLC). Subsequently, all diamonds were annealed in a vacuum chamber at 900 °C for 4 h and then boiled for three hours in an acid mixture (1 : 1 : 1 – $H_2SO_4$ : $HClO_4$ : $HNO_3$). The resulting NV centers are expected to be ~7 nm deep[35]. Photoluminescence measurements in a typical confocal setup for NV centers[36] confirmed the successful formation of NV centers with an average density of 1 NV/µm².

We measured coherence times on single NV centers in every sample by a Hahn echo measurement in a magnetic field of 400 – 600 G along the NV axis. In order to check for a potential improvement by thermal oxidation[5,19], we performed each measurement twice, once after acid cleaning of the diamond and once after an additional annealing step at 520 °C in air atmosphere. This annealing step has been calibrated beforehand on sample Scf2 as will be described below. Both Hahn echo measurements were performed on the same NV centers.

Measurements were performed in a microscope with widefield illumination and camera detection, allowing us to simultaneously probe all ~100 NV centers in an area of 10 x 10 µm² (Camera: Princeton ProEM-HS: 512B_eXcelon3, laser power: 500mW) The widefield beam was shaped from a Gaussian profile into a homogeneously illuminated square profile by a holographic phase plate TOPAG GTH-4-2.2-532.

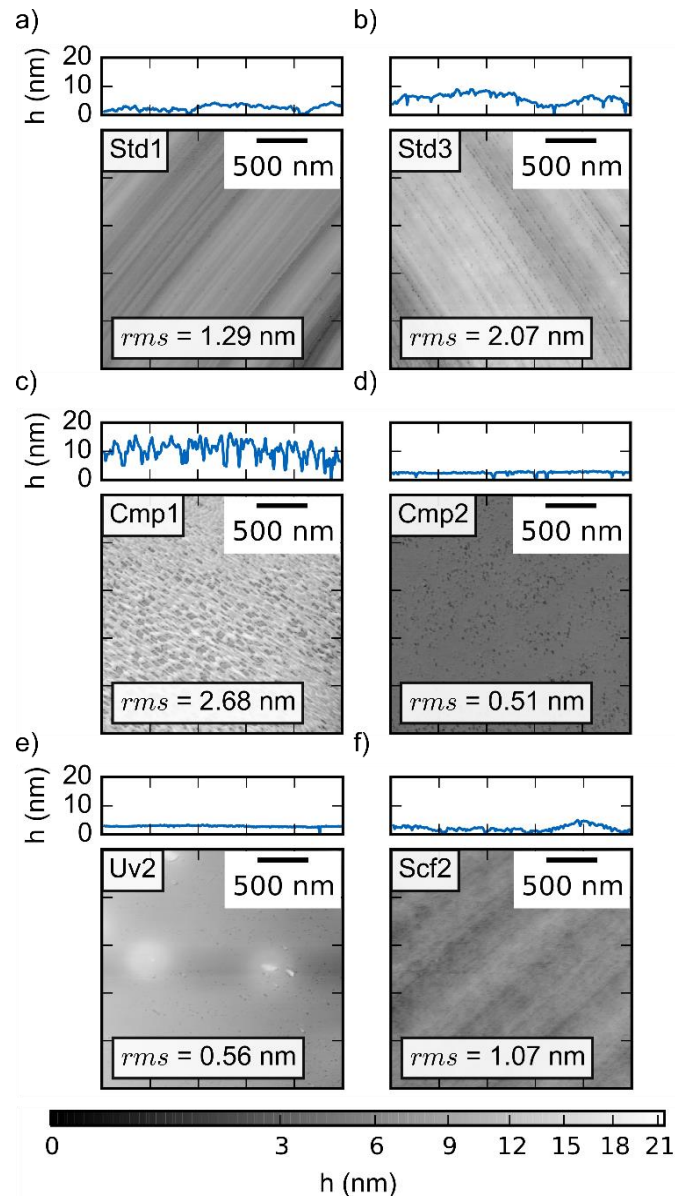

*Figure 2: AFM images of a selection of the investigated diamonds. Each scan shows a sector of 2.5 x 2.5 µm² within the area of the probed NV centers. The corresponding roughness is given in the bottom inlay. Exemplary profiles, which were chosen not to cross non-representative irregularities such as polishing debris, are given above each image. Both, profiles and images, had their lowest point set to zero for better comparability. The non-linear color bar at the bottom applies to every image. (1 column)*

## Results and discussion

In order to find proper parameters for the thermal oxidation, four treatment steps were successively applied to diamond Scf2 after the vacuum anneal: acid cleaning as described above, 4 h thermal oxidation at 465 °C in a pure oxygen atmosphere[5], 1 h thermal oxidation at 520 °C and finally 1 h thermal oxidation at 560 °C in air. After a further oxidation step at 580 °C in air, the fluorescence signal of some NV centers was quenched and the remaining NV centers showed pronounced blinking, prohibiting continued Hahn echo measurements. The samples were placed in a Piranha solution (3:1 mixture of concentrated $H_2SO_4$ and 30% $H_2O_2$) at ≤ 100 °C for 30 min immediately before and after each treatment. The distribution of $T_2$ times after each treatment is shown in Figure 3.

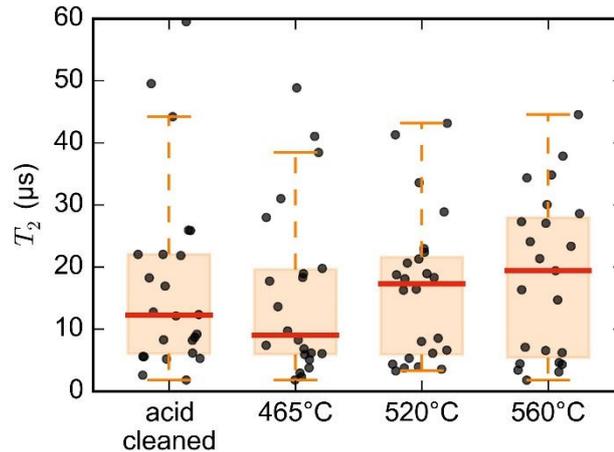

*Figure 3: Optimization of oxidation parameters. This image shows a box plot of the measured $T_2$ times of NV centers in diamond Scf2. The coherence times were measured after each surface treatment: acid cleaning, 4 h at 465 °C in oxygen, 1 h at 520 °C and 560 °C in air. Each black dot corresponds to the $T_2$ time of a single NV center and the ensemble of examined NV centers was the same for every measurement. Each dot has been displaced by a small random offset along the abscissa to improve visibility. The bottom of each box indicates the first quartile, the top indicates the third quartile and the red line in between is the median value. The whiskers show the minimum and maximum value within 1.5 interquartile range from the box. (1 column)*

While the median value, as a measure of the collective properties of the examined NV centers, could be increased from initially 12 µs to 19 µs after the 560 °C step, the maximum values decreased from 60 µs to 45 µs. Since the results from the 520 °C treatment are very similar to those of the 560 °C treatment and in order to prevent damage from too aggressive oxidation[19], we chose the 520 °C treatment as a standard for the following measurements.

We now turn to the discussion of the measured $T_2$ times for all 10 diamond surfaces. After acid cleaning (Figure 4, top), all samples display very similar coherence times. The median values are ranging from 7 µs (sample Cmp1) to 15 µs (sample Std3). Every sample shows at least one NV center with $T_2 > 40$ µs, all but three samples even have at least one NV center with $T_2 \geq 50$ µs.

As the most salient conclusion, the coherence times are surprisingly independent of the surface finish: While the shortest coherence times are found in the two roughest samples, Cmp1 and Scf1, there are also very smooth samples with comparably short coherence times, such as sample Std1 and Std2 and rather rough samples with long coherence times, such as sample Std3.

H-termination before annealing (sample H1) fails to mimic the positive effect of boron doping[28]. In addition, we observe a negative influence on the stability of photoluminescence, presumably owing to a remaining fraction of H-terminations after acid cleaning[37,38]. Roughly 60% of the NV centers were blinking strongly and therefore were not suitable for Hahn echo measurements.

The conversion yield from implanted nitrogen into NV centers appears to be independent of the surface treatment. On all samples, Hahn echo could be measured on roughly 30 centers in the measurement area. As the only exception, sample Std3 features more than twice as many data points. This, however, is due to a preferential orientation of the NV centers rather than a higher yield, resulting in more centers being properly aligned for Hahn echo measurements. We attribute this to an intrinsic property of the substrate, since Std3 has been the only sample from a different supplier.

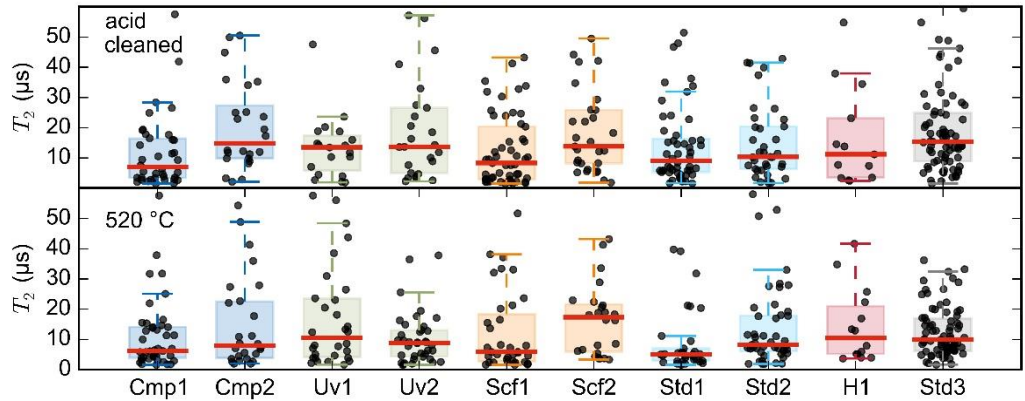

*Figure 4: Comparison of coherence times for different surface finishes and oxidation treatments. The $T_2$ measurements for each diamond are shown in box plots as introduced in Figure 3. Every black dot corresponds to the $T_2$ time of a single NV center, every red line corresponds to the median $T_2$ value. Top panel: $T_2$ measurements of all diamonds after an acid cleaning process. The resulting median $T_2$ times lie between 7 µs and 15 µs. Bottom panel: $T_2$ measurement of the same NV centers (except for sample Cmp2) after an additional step of thermal oxidation at 520 °C. The resulting median $T_2$ times lie between 5 µs and 17 µs. (2 columns)*

In a last experimental step, the $T_2$ times of the very same NV centers were measured after thermal oxidation at 520 °C for one hour in air (Figure 4, bottom). Only for sample Cmp2, where the same spot could not be found, a new spot with new NV centers had to be chosen.

Again, no systematic coherence time enhancement could be observed: The new medians are ranging from 5 µs (sample Std1) to 17 µs (sample Scf2). Although the range slightly increased, all medians, apart from the one of sample Scf2, decreased slightly, resulting in a mean median shift from 12 µs before the oxygen treatment to 9 µs afterwards, in contrast to the expected coherence time rise[19]. Regarding the number of NV centers with high coherence times (> 30 µs), no apparent trend is discernible: For some samples, the number is significantly decreased (Uv2, Std3), for others it is increased (Uv1, Std2) and some remain basically unaltered (Cmp1, Scf2).

A strong positive effect of the oxygen surface treatment was seen in the photoluminescence stability of the NV centers of sample H1. While the acid cleaning presumably left some remaining H-termination, oxygen termination by thermal oxidation appears to be more efficient. This results in a higher number of observable NV centers, comparable to the other samples. For comparison reasons, only those NVs measurable before oxidation are shown in the plot.

## Conclusion

In summary, we can answer the question raised in the title: no, ultraprecision polishing techniques cannot improve the coherence times of nitrogen-vacancy centers in diamond, at least not without additional measures. We find NV spin coherence times to be remarkably independent of the surface finish, despite the fact that roughness varies by an order of magnitude across the samples employed in this study. This fits into an emerging picture that intrinsic defects close to the surface could play a larger role in decoherence than the properties of the surface itself[28]. Accordingly, an interesting extension of this study could consist in additionally removing deep subsurface polishing defects, for instance by a plasma polish[39].

Surprisingly, a previously reported positive effect of the oxidation procedure[19] could not be reproduced. On the contrary, we even found a slight decrease of the mean median of the collective coherence times from 12 µs to 9 µs. The effect on NVs with high coherence times was unpredictable, and on average non-existing. However, thermal oxidation was found to have a positive influence on the photoluminescence of NV centers, being able to remove a formerly applied H-termination.

## Acknowledgements

The authors acknowledge support from the Deutsche Forschungsgemeinschaft (Emmy Noether grant RE3606/1-1, priority program SPP1601 and the excellence cluster NIM). We thank Patrick Simon for hydrogen terminating one sample of this study.

## Bibliography

1. Staudacher, T. *et al.* Nuclear Magnetic Resonance Spectroscopy on a (5-Nanometer)$^3$ Sample Volume. *Science* **339,** 561–563 (2013).

2. Mamin, H. J. *et al.* Nanoscale Nuclear Magnetic Resonance with a Nitrogen-Vacancy Spin Sensor. *Science* **339,** 557–560 (2013).

3. Kong, X. *et al.* Atomic-scale structure analysis of a molecule at a (6-nanometer)$^3$ ice crystal. *ArXiv170509201 Phys. Physicsquant-Ph* (2017).

4. Lovchinsky, I. *et al.* Magnetic resonance spectroscopy of an atomically thin material using a single-spin qubit. *Science* **355,** 503–507 (2017).

5. Lovchinsky, I. *et al.* Nuclear magnetic resonance detection and spectroscopy of single proteins using quantum logic. *Science* **351,** 836–841 (2016).

6. Gross, I. *et al.* Real-space imaging of non-collinear antiferromagnetic order with a single-spin magnetometer. *Nature* **549,** nature23656 (2017).

7. Kucsko, G. *et al.* Nanometre-scale thermometry in a living cell. *Nature* **500,** nature12373 (2013).

8. Fedotov, I. V. *et al.* Fiber-based thermometry using optically detected magnetic resonance. *Appl. Phys. Lett.* **105,** 261109 (2014).

9. Loretz, M. *et al.* Nanoscale nuclear magnetic resonance with a 1.9-nm-deep nitrogen-vacancy sensor. *Appl. Phys. Lett.* **104,** 033102 (2014).

10. Ofori-Okai, B. K. *et al.* Spin properties of very shallow nitrogen vacancy defects in diamond. *Phys. Rev. B* **86,** 081406 (2012).

11. Myers, B. A. *et al.* Double-Quantum Spin-Relaxation Limits to Coherence of Near-Surface Nitrogen-Vacancy Centers. *Phys. Rev. Lett.* **118,** 197201 (2017).

12. Kim, M. *et al.* Decoherence of Near-Surface Nitrogen-Vacancy Centers Due to Electric Field Noise. *Phys. Rev. Lett.* **115,** 087602 (2015).

13. Kim, M. *et al.* Effect of oxygen plasma and thermal oxidation on shallow nitrogen-vacancy centers in diamond. *Appl. Phys. Lett.* **105,** 042406 (2014).

14. Lange, G. de *et al.* Universal Dynamical Decoupling of a Single Solid-State Spin from a Spin Bath. *Science* **330,** 60–63 (2010).

15. Myers, B. A. *et al.* Probing Surface Noise with Depth-Calibrated Spins in Diamond. *Phys. Rev. Lett.* **113,** 027602 (2014).

16. Romach, Y. *et al.* Spectroscopy of Surface-Induced Noise Using Shallow Spins in Diamond. *Phys. Rev. Lett.* **114,** 017601 (2015).

17. Kim, M. *et al.* Decoherence of Near-Surface Nitrogen-Vacancy Centers Due to Electric Field Noise. *Phys. Rev. Lett.* **115,** 087602 (2015).

18. Loretz, M. *et al.* Single-proton spin detection by diamond magnetometry. *Science* 1259464 (2014).

19. Kim, M. *et al.* Effect of oxygen plasma and thermal oxidation on shallow nitrogen-vacancy centers in diamond. *Appl. Phys. Lett.* **105,** 042406 (2014).

20. Fávaro de Oliveira, F. *et al.* Effect of low-damage inductively coupled plasma on shallow nitrogen-vacancy centers in diamond. *Appl. Phys. Lett.* **107,** 073107 (2015).

21. Rosskopf, T. *et al.* Investigation of Surface Magnetic Noise by Shallow Spins in Diamond. *Phys. Rev. Lett.* **112,** 147602 (2014).

22. Stacey, A. *et al.* Nitrogen Terminated Diamond. *Adv. Mater. Interfaces* **2** (2015).

23. Chandran, M. *et al.* Nitrogen termination of single crystal (100) diamond surface by radio frequency N2 plasma process: An in-situ x-ray photoemission spectroscopy and secondary electron emission studies. *Appl. Phys. Lett.* **107,** 111602 (2015).

24. Chou, J.-P. *et al.* Nitrogen-Terminated Diamond (111) Surface for Room-Temperature Quantum Sensing and Simulation. *Nano Lett.* **17,** 2294–2298 (2017).

25. Ohno, K. *et al.* Engineering shallow spins in diamond with nitrogen delta-doping. *Appl. Phys. Lett.* **101,** 082413 (2012).

26. Osterkamp, C. *et al.* Stabilizing shallow color centers in diamond created by nitrogen delta-doping using SF6 plasma treatment. *Appl. Phys. Lett.* **106,** 113109 (2015).

27. Ito, K. *et al.* Nitrogen-vacancy centers created by N+ ion implantation through screening SiO2 layers on diamond. *Appl. Phys. Lett.* **110,** 213105 (2017).

28. Fávaro de Oliveira, F. *et al.* Tailoring spin defects in diamond by lattice charging. *Nat. Commun.* **8,** (2017).

29. Schuelke, T. & Grotjohn, T. A. Diamond polishing. *Diam. Relat. Mater.* **32,** 17–26 (2013).

30. Thomas, E. L. H. *et al.* O. A. Silica based polishing of {100} and {111} single crystal diamond. *Sci. Technol. Adv. Mater.* **15,** (2014).

31. Watanabe, J. *et al.* Ultraviolet-irradiated precision polishing of diamond and its related materials. *Diam. Relat. Mater.* **39,** 14–19 (2013).

32. Kubota, A. *et al.* Development of an Ultra-finishing technique for single-crystal diamond substrate utilizing an iron tool in H2O2 solution. *Diam. Relat. Mater.* **64,** 177–183 (2016).

33. Hahn, E. L. Spin Echoes. *Phys. Rev.* **80,** 580–594 (1950).

34. Thomas, E. L. H. *et al.* Chemical mechanical polishing of thin film diamond. *Carbon* **68,** 473–479 (2014).

35. Pezzagna, S. *et al.* Creation efficiency of nitrogen-vacancy centres in diamond. *New J. Phys.* **12,** 065017 (2010).

36. Gruber, A. *et al.* Scanning Confocal Optical Microscopy and Magnetic Resonance on Single Defect Centers. *Science* **276,** 2012–2014 (1997).

37. Hauf, M. V. *et al.* Chemical control of the charge state of nitrogen-vacancy centers in diamond. *Phys. Rev. B* **83,** 081304 (2011).

38. Seshan, V. *et al.* Hydrogen termination of CVD diamond films by high-temperature annealing at atmospheric pressure. *J. Chem. Phys.* **138,** 234707 (2013).

39. Ding, G. F. *et al.* Micromachining of CVD diamond by RIE for MEMS applications. *Diam. Relat. Mater.* **14,** 1543–1548 (2005).